\documentclass[prl,twocolumn,showpacs,preprintnumbers,amsmath,amssymb]{revtex4}

\usepackage{graphicx}
\usepackage{dcolumn}
\usepackage{bm}

\begin{document}


\title{Topological defects in two-dimensional crystals}

\author{Yong Chen}
\altaffiliation{Email: ychen@gmail.com}
\affiliation{Institute of Theoretical Physics, Lanzhou University, Lanzhou $730000$, China}

\author{Wei-Kai Qi}
\altaffiliation{Email: weikaiqi@gmail.com}
\affiliation{Institute of Theoretical Physics, Lanzhou University, Lanzhou $730000$, China}

\date{\today}

\begin{abstract}
By using topological current theory, we study the inner topological structure of the topological defects in two-dimensional (2D) crystal. We find that there are two elementary point defects topological current in two-dimensional crystal, one for dislocations and the other for disclinations. The topological quantization and evolution of topological defects in two-dimensional crystals are discussed. Finally, We compare our theory with Brownian-dynamics simulations in 2D Yukawa systems.
\end{abstract}


\maketitle

In 1970's, Kosterlitz and Thouless construct a detailed and complete theory of superfluidity on two-dimensions (2D)~\cite{kt}. They indicate vortices pair unbinding will lead to a second-order transition in superfluid films. Later, a microscopic scenario of 2D melting has been posited in the form of the Kosterlitz-Thouless-Halperin-Nelson-Young (KTHNY) theory~\cite{hny}.  The KTHNY theory predicts the unbinding of topological defects to break the symmetry in the two-stage transitions. Topological defects, which are a necessary consequence of broken continuous symmetry, play a important role in two-dimensional phase transition~\cite{Dn,Cl}.

In 2D crystal, the evolution of topological defects have been studied in experiments and computer simulations. A serial experiments were performed to calculate dislocations and disclinations dynamic of two-dimensional colloidal systems, and dissociation of dislocations and disclinations were observed~\cite{td001}. During the years, a large number of computer simulations indicated that exist a two-stage melting as prescribed by KTHNY theory, however, results are still controversial~\cite{KnNs}. Our previous work found that exist a hexatic-isotropic liquid phase coexistence during the melting of soft Yukawa systems~\cite{Qi}. By Voronoi polygons analysis, the behavior of piont defects in the coexistence is very complicated. The evolution of topological defects during the melting of two-dimensional system still a open question.

It is interesting to consider the appropriate form for the point defect densities when expressed in terms of the vector order parameter field $\phi(\vec{r},t)$. This has been carried out by Halperin~\cite{Ha}, and exploited by Liu and Mazenko~\cite{LM}. However, their analysis is incomplete~\cite{Duan00}. In two-dimensional system, a gauge field-theoretic formalism developed by Kleinert~\cite{HK}. The gauge theory of topological quantum melting in 2+1 dimensions Bose system was developed by Nussinov et al, and the Superfluidity and superconductivity can arise in a strict quantum field-theoretic setting~\cite{Z1}.

Recently, a topological field theory for topological defect developed by Duan et al~\cite{Duan}. By using $\phi$-mapping method and topological current theory, the evolution of topological defect which relate to sigularities of the order-paramter field, such as vortex in BEC~\cite{Duan04} and superconductivity~\cite{Duan05}, was studied. In this Letter, we will develop a topological current theory of discloations and disclination in 2D crystals. By using the topological field approach, the topological quantization and evolution of topological defects in 2D crystals was discussed. We compared our theory with Brownian-dynamics simulations in Yukawa systems.

In continuum elasticity theory, the homogeneous equation in 2D triangular solid is given by~\cite{Landau, Nelson}
\begin{equation}
\nabla^4\chi/Y=\epsilon_{kl}\partial_{k}\partial_{l}\theta+\epsilon_{ik}\partial_{k}(\epsilon_{jl}\partial_l\partial_j u_i)
\label{eq:elasticity}
\end{equation}
where $Y=4\mu(\mu+\lambda)/(2\mu+\lambda)$ is the two-dimensional Young's modulus and $\sigma_{0}=\lambda/(2\mu+\lambda)$ the two-dimensional Poisson ratio. The defects associated with the continuum elastic theory of a solid are dislocations and disclinations. In the following we consider only the triangular lattice since it is the most densely packed one in two-dimensional and favored by Nature.

\begin{figure}
\begin{center}
\includegraphics[width=0.15\textwidth]{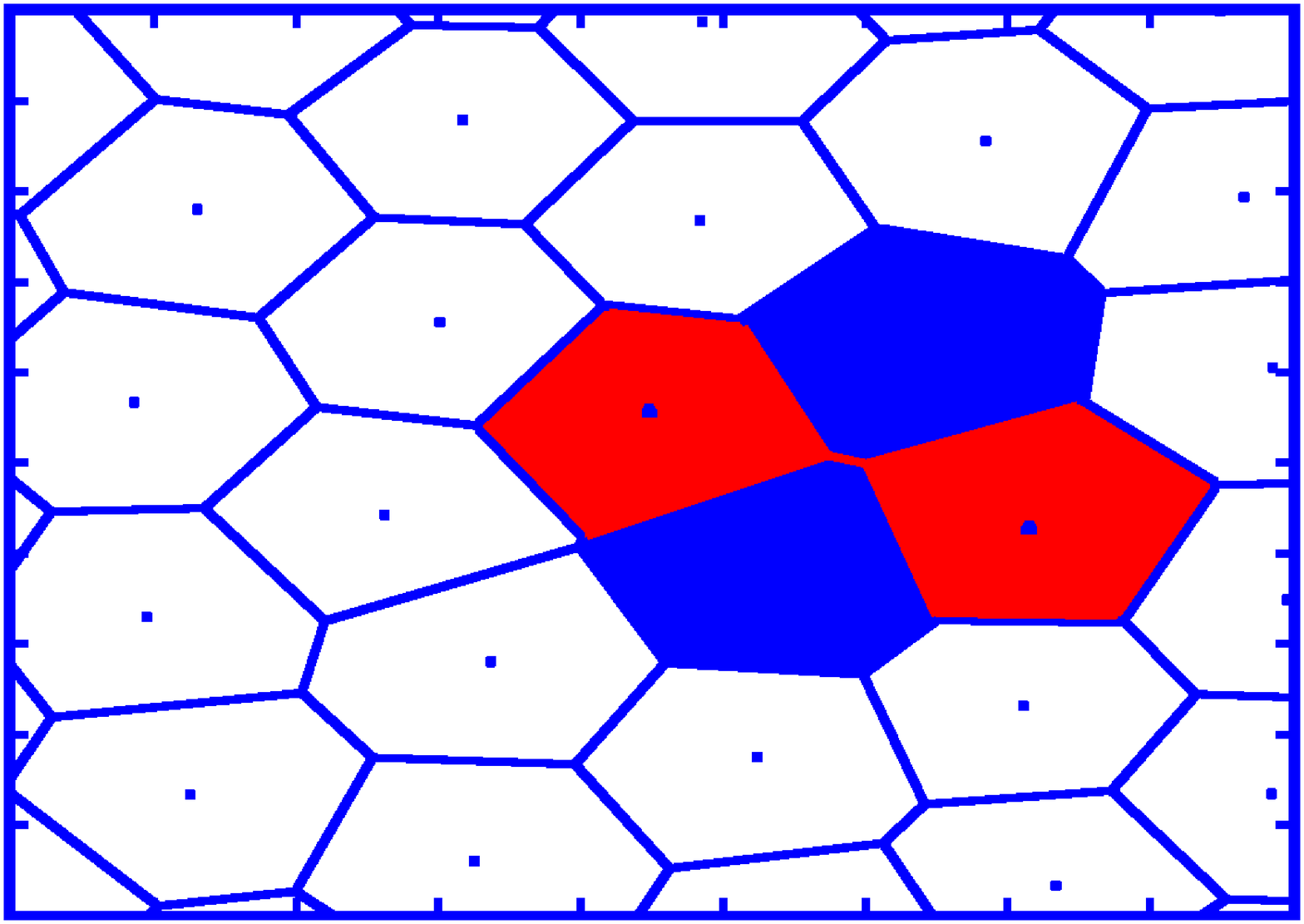}
\includegraphics[width=0.15\textwidth]{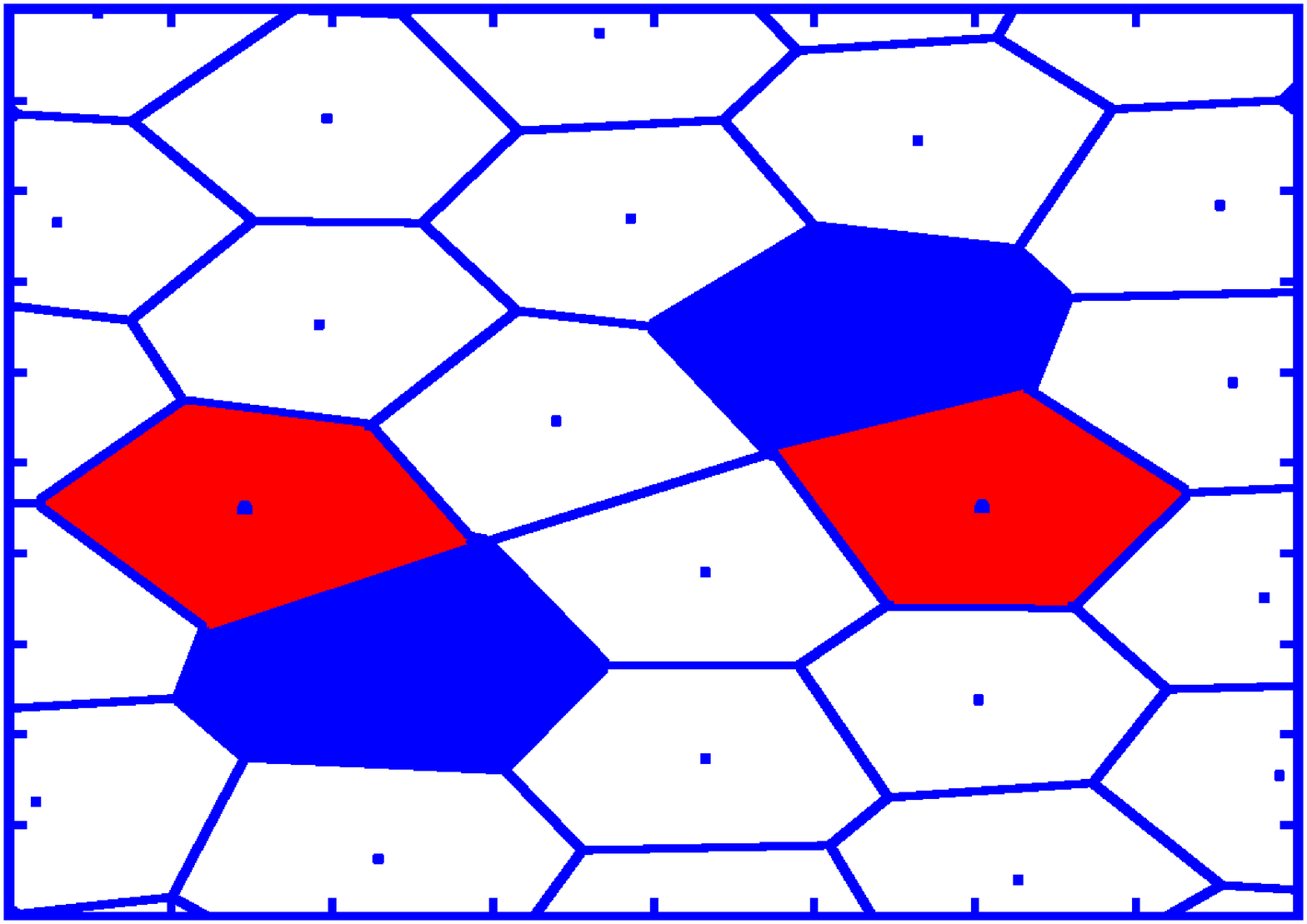}
\includegraphics[width=0.15\textwidth]{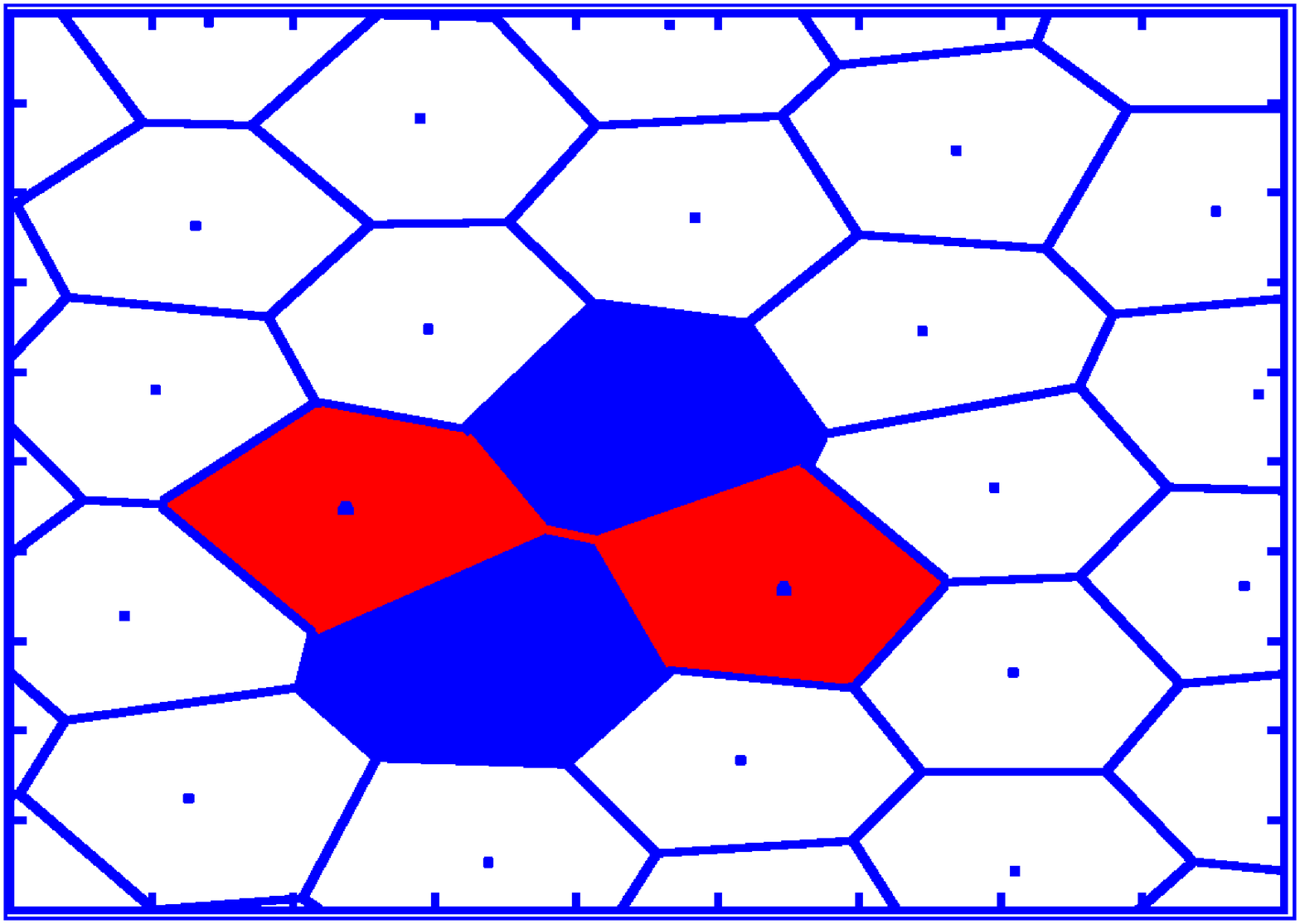}
\includegraphics[width=0.15\textwidth]{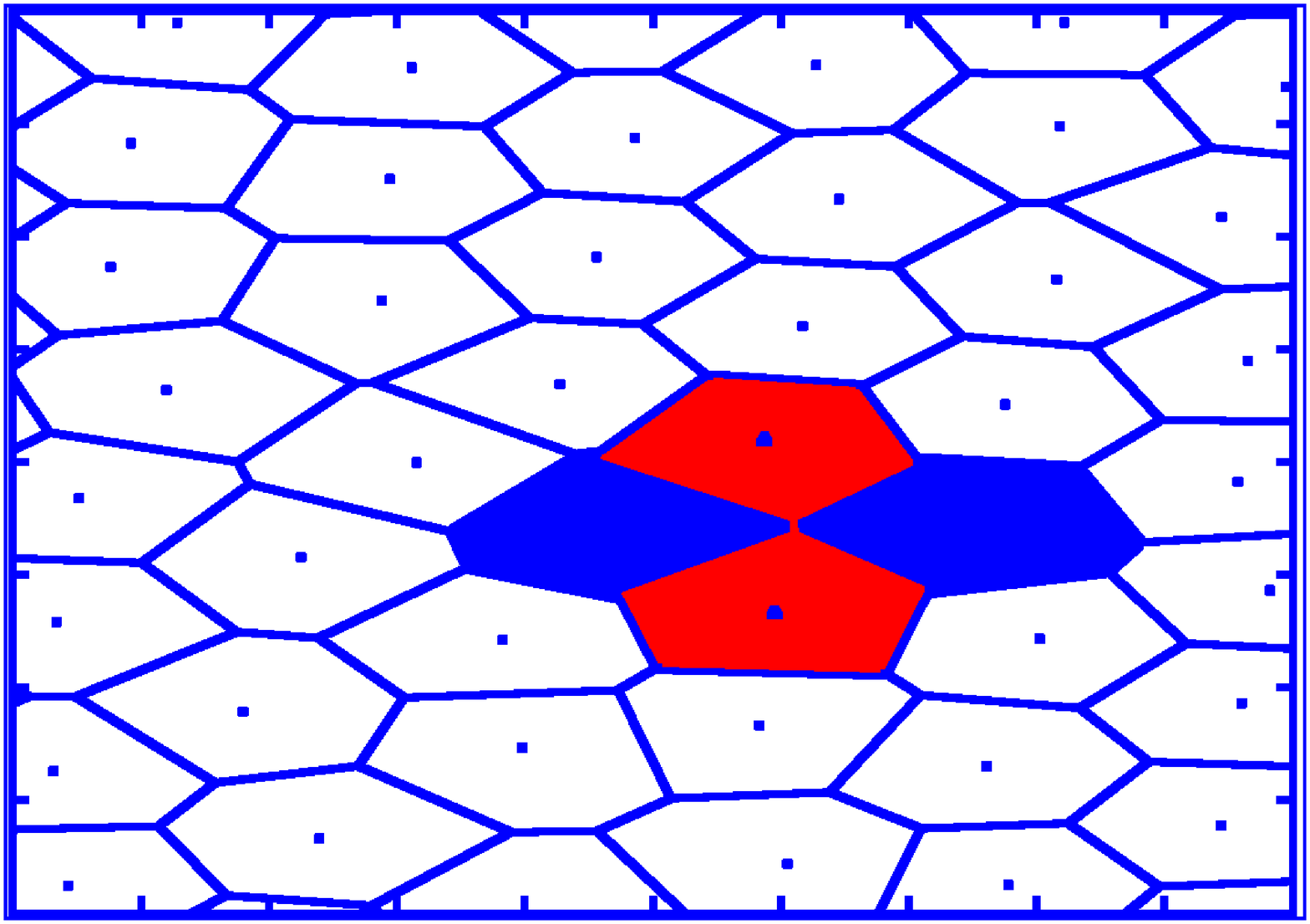}
\includegraphics[width=0.15\textwidth]{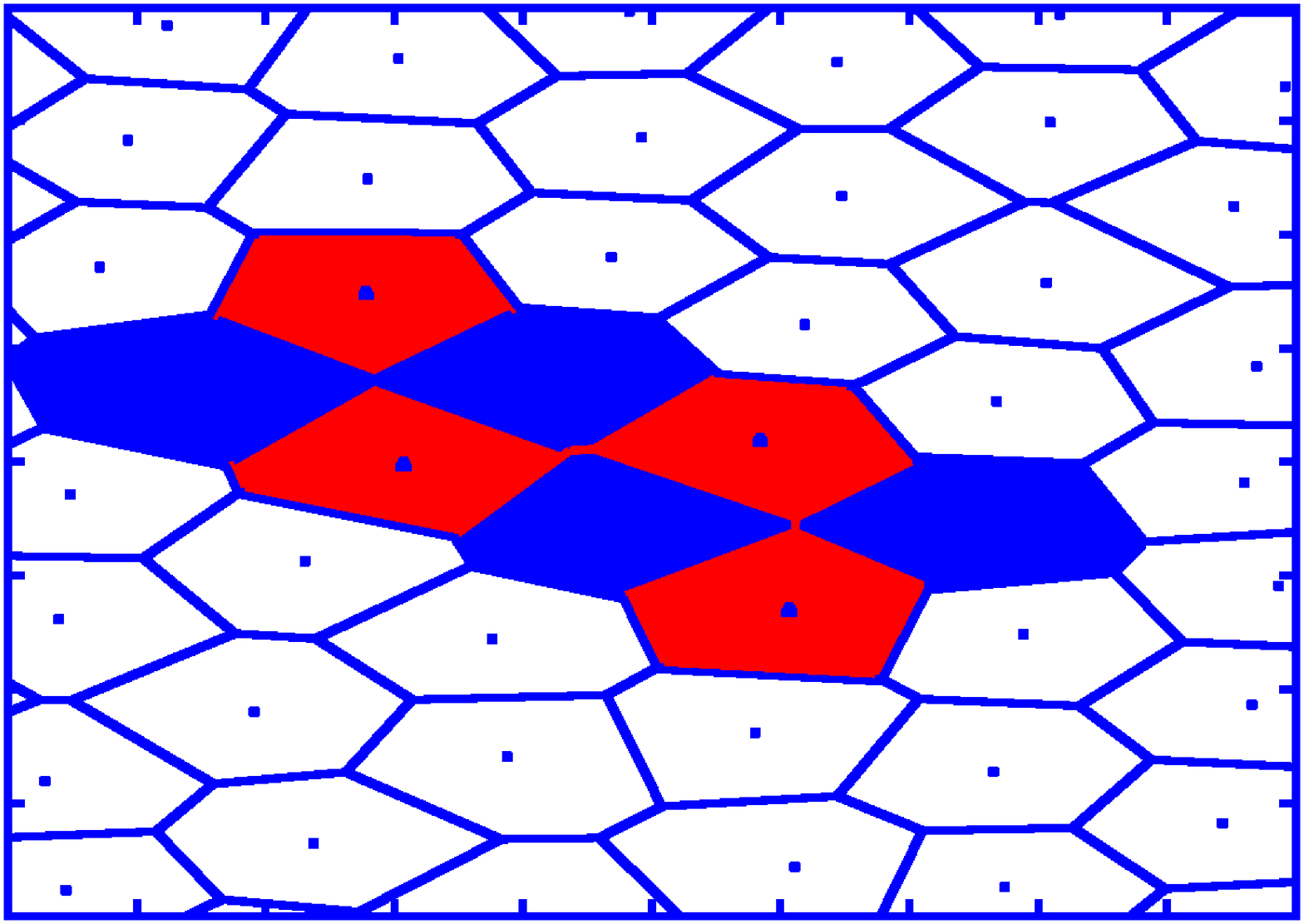}
\includegraphics[width=0.15\textwidth]{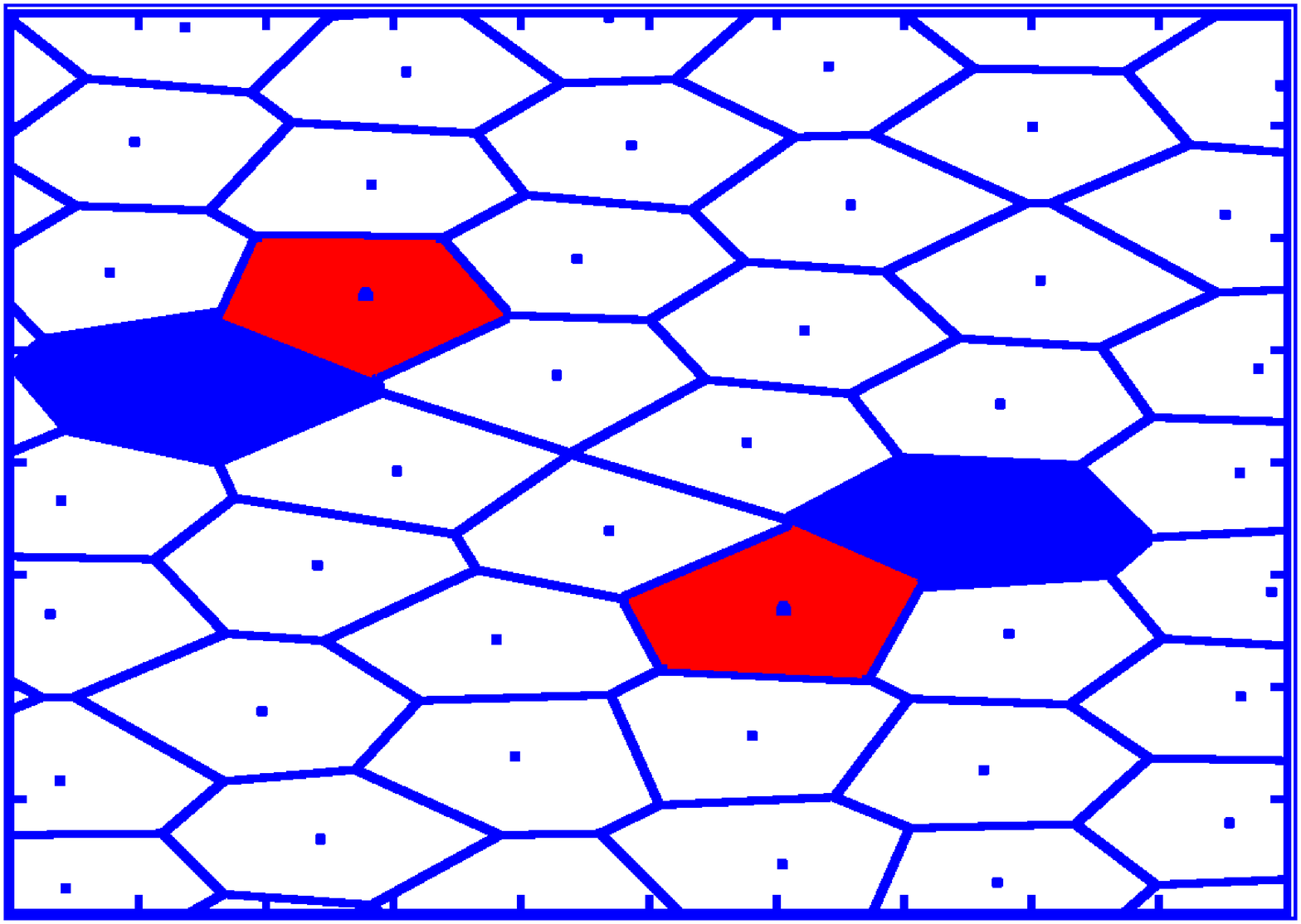}
\caption{Voronoi polygons analysis in 2D Yukawa systems.}
\label{Fig:A}
\end{center}
\end{figure}

Topological defects associated with the continuum elastic theory of a two-dimensional solid
are dislocations and disclinations (see Fig.~\ref{Fig:A}). Disclinations can be introduced into the theory in a way similar to the discussion of vortices in two-dimensional XY model~\cite{QY}. Since the bond angle is defined only up to $2\pi/6$, it implies
\begin{equation}
\oint d\theta=\oint\nabla\theta d\vec{s}=\frac{2\pi}{6}W^6, \quad W^6=0, \pm1,\ldots,n.
\label{eq:Topdisclination}
\end{equation}
Where $W^6$ is the winding number of disclinations. Disclinations, which can influence on the decay of orientational order, have a much higher energy than dislocations. They are defined in terms of the bond angle field $\theta(r)\equiv1/2(\partial u_y/\partial x-\partial u_x/\partial y),$
which measures the bonds orientational order. It is convenient to define an order parameter for bond orientations, which for the triangular lattices is $\psi(r)=\psi_0e^{i6\theta(r)}$. However, the bond angle field is undefined at the disclination cores, i.e., the zero points of the order parameter. We rewrite the orientation order parameter $\psi(r)=\phi_6^1+i\phi_6^2$ instead of $\psi(r)=\psi_0e^{i6\theta(r)}$. We can define the unit vector field $\vec{n}_6(\textbf{x})$ as $n_6^a=\phi_6^a/||\phi_6||$ and $||\phi_6||=\sqrt{\phi_6^a\phi_6^a}$, $a=1, 2$. We can construct a topological current of the orientation order parameter field in two-dimensional crystal, which carries the topological information of $\vec{\phi}(x)$,
\begin{equation}
j_{disc}^k=\frac{1}{6}\epsilon^{ijk}\epsilon_{ab}\partial_i n_6^a\partial_j
n_6^b.
\label{Curdisclination}
\end{equation}
It is the $\phi$-mapping topological current for disclinations. We will see that this current associate with the singularities of the unit field $\vec{n}_6(x)$. It is easy to see that this topological current is identically conserved, i.e., $\partial_kj_{disc}^k=0$.

We consider only the triangular lattice since it is the most densely packed one in
two-dimensional and favored by Nature. The dislocation is characterized by a Burgers vector $\vec{b}$, which can be determined by considering a loop enclosing the dislocation. In mathematical language,
\begin{equation}
\oint du_l=\oint \partial_i u_l dx_i=b_l.
\label{eq:Topdislocation}
\end{equation}
By introduce two-dimensional local transition order parameters as
$\rho_x=\rho_{0x}e^{i G_x u_x}=\phi_x^1+i\phi_x^2$ and $\rho_y=\rho_{0y}e^{i G_y u_y}=\phi_y^1+i\phi_y^2$,
where $\textbf{G}$ is any reciprocal lattice vector and $\rho_0=e^{i G_0 r_0}$. We can define a unit vector field $\vec{n}_l(x)$, where
$\vec{n}_l(\textbf{x})=e^{i G_l u_l}=n_{l}^1+i n_{l}^2$ ($l=x, y$).
It can be further expressed as
\begin{equation}
n_l^a(\textbf{x})=\frac{\phi_l^a(\textbf{x})}{||\phi_l(\textbf{x})||}, \quad ||\phi_l(\textbf{x})||=\sqrt{\phi_l^a(\textbf{x})\phi_l^a(\textbf{x})},
\nonumber
\end{equation}
obviously, $n_l^an_l^a=1$. We can construct topological currents of the order parameter field in two-dimensional crystal, which carries the topological information of $\vec{\phi}_{l}(x)$,
\begin{equation}
j_{diso}^{kl}=\frac{1}{G_l}\epsilon^{ijk}\epsilon_{ab}\partial_i n_l^a\partial_j
n_l^b.
\label{eq:Curdislocation}
\end{equation}
where the Burgers vectors satisfy $b_lG_l=2\pi$. It is the $\phi$-mapping current for dislocations. Dislocations can be quite effective at breaking up translational order. However, they are less disruptive of orientational correlations. The disclination can be seen as a bound of dislocations with parallel Burger vectors.  On the hand,
disclinations, which play a important role in the disruptive of the orientational order in two-dimensional solid, can be seen as bound disclination and anti-disclination pairs.

From the relationship between the bond angle field and displacement vector field that $\partial_i\theta=-\epsilon_{ab}\partial_a\partial_i(u^b)$. Using the topological current theory, we can obtain
\begin{equation}
\partial_{k}j_{diso}^{kl}=-\epsilon_{li}j_{disc}^{i},
\label{eq:Topsource}
\end{equation}
it indicate that disclinations can be seen as sources for dislocations. Since there aren't free disclinations in two-dimensional crystal at the low temperature phase, the dislocation currents are conserved as $\partial_{k}J_{k}^{l}=0$.

By using Green function relation and substitute $\phi_6^a$ to the homogeneous equation~\cite{Duan}, we can rewrite the homogeneous equation into a $\delta$-function form as
\begin{equation}
\frac{1}{Y}\nabla^4\chi=\frac{2\pi}{6}\delta^2(\vec{\phi_6})J\big(\frac{\phi_6}{x}\big)+\epsilon_{kl}\partial_k\frac{2\pi}{G_l}\delta^2(\vec{\phi_l})J\big(\frac{\phi_l}{x}\big).
\label{eq:deltacurrent}
\end{equation}

With the $\delta$-function theory, the homogeneous equation can deduce into
\begin{equation}
\frac{\nabla^4\chi}{Y}=\sum_{i}\beta_i\eta_i\delta^2(\vec{r}-\vec{z_i}(t))+\sum_{j}\beta_{j}^{l}\eta_{j}^{l}\epsilon_{kl}\partial_k\delta^2(\vec{r}-\vec{z_j}(t)),
\label{eq:deltahomo}
\end{equation}
where $\eta_i=\pm1$ is Brouwer degree, the positive integer $\beta_i$ is called the Hopf index of map $x\rightarrow\vec{\phi}$. The topological charge of disclination and dislication is $\beta_i\eta_i=2\pi/6$ and $\beta_{j}^{l}\eta_{j}^{l}=b_{j}^{l}$. The first term is related to dislocation, and the second term is related to disclination. Also, it can simply seen as disclination current, if dislocations are thought of as disclination dipole pairs.

We explore what will happen to the disclination at the zero point $(t^*, \vec{z_l})$. According to the values of the vector Jacobian at zero points of the order parameter, there are limit points ($J^0(\phi_6/x)\neq0$) and bifurcation points ($J^0(\phi_6/x)=0$). Each kind corresponds to different cases of branch processes. The limit points are determined by
\begin{equation}
\left. J^{0} \left( \frac{\phi_6}{x} \right) \right|_{t^*, \vec{z_l}} = 0, \quad
\left. J^{1} \left( \frac{\phi_6}{x} \right) \right|_{t^*, \vec{z_l}} \neq 0.
\label{eq:condition01}
\end{equation}
Considering the condition (\ref{eq:condition01}) and making use of the implicit function theorem, the solution in the neighborhood of the point ($t^*, \vec{z_l}$) is $t=t(x),~y=y(x)$, where $t^*=t(z_l^1)$. In this case, one can see that $ \left. dx/dt \right|_{(t^*,\vec{z_l})}=\infty$, the Taylor expansion of $t=t(x)$ at the limit points $(t^*,
\vec{z_l})$ is
\begin{equation}
t-t^* = \left. \frac{1}{2} \frac{d^2t}{dx^2} \right|_{t^*, \vec{z_l}} (x-z_l^1)^2,
\label{eq:xt01}
\end{equation}
which is a parabola in the x-t plane. From this equation, we can obtain two solutions $x_1(t)$ and $x_2(t)$, which give two branch solutions (World lines of disclinations). If $d^2t/dx^2|_{(t^*, \vec{z_l})}>0,$ we have the branch solutions for $t > t^*$. It is related to the origin of a disclination pair. Otherwise, we have the branch solutions for $t < t^*$, which related to the annihilation of a disclination pair. Since the topological current is identically conserved, the topological charges of these two generated or annihilated
disclinations must be opposite at the limit points, say $\beta_1\eta_1+\beta_2\eta_2=0$.
For a limit point it is required that $J^1(\phi/x)|_{t^*, \vec{z_l}}\neq0$.

At the neighborhood of the limit point,one can obtain the scaling law as~\cite{Xu}
$v\varpropto(t-t^{*})^{-1/2}$. The growth or annihilation is parameterized in terms of a relevant characteristic length $\xi(t)$, which is also characteristic the mean separation distance of disclinations pair. The relevant length $\xi(t)$ obeys that
$\xi(t)\sim (t-t^{*})^{1/2}$. It is the dynamic scaling law of the topological defects pairs. In the low temperature equilibrium phase has essentially no disclination pairs, and $\xi(t)$ is infinite. The relationship between the relevant length and the metastable disclination density which below the critical temperature is$\xi(t)=\sqrt{1/\rho_v}$.
then the number of disclinations satisfy the power law $N\propto t^{-1}$.

Consider in which the restrictions on zero point $(t^*, \vec{z_l})$ are $J^{k}(\phi_6/x)\big|_{(t^*, \vec{z_l})}=0, (k=0,1,2)$, which imply an important fact that the function relationship between t and x or y is not unique in the neighborhood of the bifurcation point $(t^*, \vec{z_l})$. With the aim of finding the different directions of all branch curves at the bifurcation point, we suppose $\partial \phi_6^1/\partial y\big|_{t^*, \vec{z_l}}\neq0$. According to the $\phi$-mapping theory, the Taylor expansion of the
solution of the zeros of the order parameter field in the neighborhood of $(t^*,
\vec{z_l})$ can be expressed as
\begin{equation}
A(x-z_l^1)^2 + 2B(x-z_l^1)(t-t^*) + C(t-t^*)^2 + \cdots = 0,
\label{eq:xttwo}
\end{equation}
where A, B, and C are constants determined by the order parameter. The solutions of Eq.~(\ref{eq:xttwo}) give different directions of the branch curves (world line of disclinations) at the bifurcation point. There are four possible cases, which will show the physical meanings of the bifurcation points.

First we consider the case that $A\neq0$. If $\Delta=4(B^2-AC)>0$, from Eq.(\ref{eq:xttwo})
we get two different motion directions of the core of disclination $dx/dt\big|_{1,2}=(-B\pm\sqrt{B^2-AC})/A$, where two topological defects encounter and the depart at the bifurcation point. However, when $\Delta=4(B^2-AC)=0$, form Eq.~(\ref{eq:xttwo}), we obtain only one motion direction of the core of disclination$dx/dt\big|_{1,2}=-B/A$, which includes three important cases. (i) Two disclinations tangentially encounter at the bifurcation point. (ii) Two disclinations merge into one disclination at the bifurcation point. (iii) One disclinations splits into two disclinations at the bifurcation point.

Then we consider the case that $A=0$, If $C\neq0$ and $\Delta=4(B^2-AC)=0$, from Eq.~(\ref{eq:xttwo}) we have $dt/dx\big|_{1,2}=(-B\pm\sqrt{B^2-AC})/C=0,-(2B)/C$. There are two important cases. (i) One disclination split into three disclinations at the bifurcation point. (ii)Three disclinations merge into one disclination at the bifurcation point. From the above discuss, we can then obtain the growth or annihilation velocity of the vortices as $v \propto const$. The approximation asymptotic relation of  is $\xi(t) \propto(t-t^*)$. If $C=0$, Equations (\ref{eq:xttwo}) give $dx/dt=0$ or $dt/dx=0$. This case shows that two worldlines intersect normally at the bifurcation point, which is similar to the case that $A=0$ and $C\neq0$. We can obtain $\xi(t) = const,~ v= 0$.

It is obvious that vortices are relatively at rest when $\xi(t)=const$. Through our topological current theory, we can get the dynamical scaling law. Moreover, it worth to emphasizing that the dynamical scaling law of vortices, only depend on topological properties of the order parameter field.

\begin{figure}
\begin{center}
\includegraphics[width=0.5\textwidth]{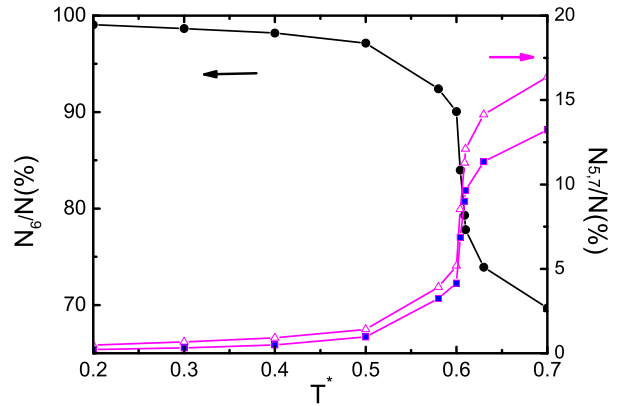}
\caption{(a)The fractions of $6$-coordinated, $5$-coordinated, and
$7$-coordinated particles. At low temperature the number of defects
is very small. When the temperature above $0.500$, $N_{6}/N$
decreases rapidly.}
\label{Fig:B}
\end{center}
\end{figure}

We performed Brownian dynamics simulations on melting of two-dimensional colloidal crystal in which particles interact with Yukawa potential~\cite{Qi}. We have observed the mechanisms of defects in the two-dimensional Yukawa system~\cite{mv}. Fig.~\ref{Fig:B} plots the fractions of $6$-coordinated, $5$-coordinated, and $7$-coordinated particles. At low temperature, all particles are nearly six-coordinated, and the number of defects
is very small. When the temperature reaches $0.500$, $N_{6}/N$ behaves in a rapidly decreasing. At $T^{*}=6.05$, almost $20\%$ of particles are attached to the defects, which consisted with the results in Ref.~\cite{Ta}. As the defects fraction rises above
$30\%$, the system melt into a liquid phase.  At the low temperature, the paired dislocation is formed or annihilated. With the temperature rises to the hexatic phase, the dislocation pair comes into dissociated. In the solid phase, the unbinding of dislocation is unstable, and these unbind dislocation will bind soon, whereas the stable free dislocation is
found in the hexatic phase. The above solutions form topological current theory reveal the evolution of disclinations.

Besides the encountering of disclinations, splitting and merging of disclinations are also included. When a multicharged disclinations, such as $8$-coordinated particles, pass the bifurcation point, it may split into several disclinations with lower value of Burgers vector along different branch curves. On the other hand, several disclinations can merge into one disclination at the bifurcation point. As before, since the topological current of disclinations is identically conserved, the sum of the topological charge of final disclinations must be equal to that the initial vortices at the bifurcation point, i.e. $\sum_f\beta_{lf}\eta_{lf}=\sum_i\beta_{li}\eta_{li}$ for fixed l. It indicates that, vortices with a higher value of Burgers vector can evolve to the lower value of Burgers vector, or vortices with a lower value of Burgers vector can evolve to the higher value of Burgers vector through the bifurcation process. But due to  Such case was observed in our BD simulation in Yukawa systems. That is why at the liquid phase the number of $5$-coordinated particles is much more than $7$-coordinated particles due to the emergence of the $8$-coordinated particles. The similar experimental result was observed in melting of two-dimensional tunable-diameter colloidal crystals~\cite{Han}.

\bigskip

\begin{acknowledgments}
We would like to thank Y. X. Liu for his helpful discussions. This work was supported by  the National Natural Science Foundation of China and by the SRF for ROCS, SEM.
\end{acknowledgments}

\end{document}